\documentclass[twocolumn, prl]{revtex4} 
\usepackage{amsmath}
\usepackage{bm}

\def \pd {\partial}
\def \v#1{{\bm #1}}
\begin{document}

\title{Model for Motions of Impurities in Bose-Einstein Condensates}
\author{Jun Suzuki$^*$}
\affiliation{Department of Physics and Astronomy, University of South Carolina, Columbia, SC 29208}
\date{January 21, 2005}

\begin{abstract}
A model for classical impurities moving in Bose-Einstein 
Condensate (BEC) is proposed in the framework of quantum field 
theory and is solved within the Bogoliubov approximation at zero 
temperature. Several formulae are obtained for physical quantities 
such as the occupation number, the dissipation, the free energy, 
and the depletion. To illustrate this model two examples are 
studied; an impurity moving with i) a constant velocity and ii) a 
constant acceleration. Landau's criterion for energy dissipation 
accompanying the motion of impurities due to creation of 
quasi-particles in a superfluid, which was proposed on the basis 
of purely phenomenological argument, is obtained in the framework 
of an exact quantum mechanical model.
\end{abstract}

\maketitle


The recent experimental discovery of BEC in alkali vapors \cite{bec} was facilitated by 
magnetic trapping and the development of novel cooling techniques. 
This discovery and subsequent experiments motivated extensive 
theoretical activity and studies of BEC with additional trapping potentials \cite{review}.

Impurities are present in all experimental situations and should be accounted for. 
Here we discuss dissipation associated with motion of impurities in BEC. 
A seminal work on the problem of motion of impurities in a superfluid is due to Landau \cite{landau,landausp2}.  
He proposed a simple criterion for energy dissipation based on 
purely phenomenological considerations demonstrating that an 
impurity moving with a velocity greater than some critical value 
generally produces excitations in a fluid. Superfluidity is 
explained then by the absence of excitations for motions with 
velocities smaller than the critical Landau velocity. 
We sketch here shortly Landau's original argument for the sake of later 
comparison with the results of our model computation of the 
effects of motion of impurities in the weakly degenerate Bose-Einstein gas. 

Suppose the excitation in a fluid, whose atoms have mass $M$, has an energy $E$ 
and a momentum $ \v p$ in a reference frame where the fluid is at rest. 
Then the energy of the excitation $E '$ in the coordinate system where the fluid is moving with velocity 
$\v v$ with respect to the container, is given, in classical mechanics, by a Galilean boost: 
$E ' = E + \v p\cdot \v v + M v^2/2$.  
Dissipation due to the motion of the excitation can happen if the energy in the boosted frame is negative, 
$ E + \v p \cdot \v v<0$.  In the optimal case of $\v p$ anti-parallel to $\v v$ this implies $v>E/ p$. 
Therefore, for the phonon-like excitation $E =p c$ this condition gives $v>c$, with $c$ the speed of sound in the fluid. 
While this condition is widely accepted, it seems that it was not derived yet from an underlying microscopic theory. 

Static ``quenched" randomly distributed non-dynamic impurities were studied in \cite{huang, kobayashi}. 
The problem of impurities moving in condensates has not been studied sufficiently 
in its full generality until relatively recent partial results \cite{meystre, kovrizhin, ms, astracharchik}. 
Dynamical quantum impurities were also studied recently \cite{mazets, mazets2}. 

The purpose of this Letter is to derive Landau's 
criterion from a microscopic quantum many body theory, 
and to study more general effects of impurities in BEC in a consistent and 
unified approach. To this end we propose a model
Hamiltonian with classical impurities moving along given
trajectories in BEC and diagonalize it using Bogoliubov's
method \cite{bogoliubov}. Landau's criterion appears naturally.
We also compute the effects of an accelerated impurity that 
can hopefully be experimentally tested. 

The Bogoliubov Hamiltonian for interacting bosons of mass $M$ is 
\begin{multline} \label{H0}
\hat{H}_0 = \int d^3 x \ [\hat{\psi}^{\dagger} (\v{ x}) \left(-\frac{\hbar ^2 \v{ \nabla ^2}}{2 M}\right)\hat{\psi} (\v{ x}) \\
+ \frac g2  \hat{\psi}^{\dagger} (\v{ x}) \hat{\psi}^{\dagger} (\v{ x}) \hat{\psi} (\v{ x}) \hat{\psi} (\v{ x})] .
\end{multline}
The coupling constant $g$ is related to the 
s-wave scattering length $a_s$ by $g=4 \pi a_s \hbar ^2/M$. 
We suggest adding the interaction Hamiltonian; 
\begin{equation} \label{Hi} 
\hat{H}_{{\rm int}} (t) =\sum _i \lambda_i  \int d^3x \ \delta (\v{ x}-\v{ \zeta}_i(t)) \hat{\psi}^{\dagger} (\v{ x}) \hat{\psi} (\v{ x}) , 
\end{equation}
to account for the local interaction between bosons and impurities moving along given trajectories $\v{ \zeta}_i(t)$. 
The coupling constants $\lambda _i$ between $i$-th impurity and bosons are related to the corresponding 
s-wave scattering lengths $b_s^i$ and reduced masses $m_i$ by $ \lambda _i=2 \pi b_s^i \hbar ^2/m_i$. 
Expanding the fields $\hat{\psi}^{\dagger} (\v{ x})$ and $\hat{\psi} (\v{ x})$ in terms of 
creation and annihilation operators in the plane wave basis in a box of volume $V=L^3 $, we obtain, 
\begin{multline} \label{H2} 
\hat{H}(t) = \sum _{\v k} \epsilon _{k} \hat{a}_{\v k}^{\dagger}    \hat{a}_{\v k}
+ \frac{g}{2V} \sum _{\v k,\v k',\v q} \hat{a}_{\v k+\v q}^{\dagger}   \hat{a}_{\v k'-\v q}^{\dagger}   \hat{a}_{\v k}   \hat{a}_{\v k'}   \\
+ \sum _i  \frac{\lambda_i}{V} \sum _{\v k,\v k'} e ^{-i(\v k-\v k')\cdot \v{\zeta}_i (t)} \hat{a}_{\v k}^{\dagger}    \hat{a}_{\v k'}   ,
\end{multline}
where $ \epsilon _{k}=\hbar ^2 \v k^2/2M$ is the free particle kinetic energy. 
In the thermodynamic limit ($N,V \to \infty$ with number density $n=N/V$ finite), 
$\lim_{V\to \infty} (1/V)\sum_{\v k}= (2\pi)^{-3} \int d^3 k$, 
$\lim_{V\to \infty} \sqrt{V}\hat{a} ^{\dagger}_{\v k}=(2\pi)^{3/2}\hat{a} ^{\dagger}(\v k)$, and 
$\lim_{V\to \infty} \sqrt{V}\hat{a} _{\v k}=(2\pi)^{3/2}\hat{a}(\v k)$.  Then 
$\hat{a} ^{\dagger}(\v k)$ and $\hat{a}(\v k)$ satisfy the usual commutation relations: 
$[\hat{a}(\v k),\;\hat{a} ^{\dagger}(\v k')]  = \delta(\v k-\v k')$, 
$\ [\hat{a}(\v k),\;\hat{a} (\v k')] =[ \hat{a} ^{\dagger}(\v k),\;\hat{a} ^{\dagger}(\v k')]=0$.

In the Bogoliubov original treatment \cite{bogoliubov} of the above 
system the total number of particles is not conserved. 
In the following we adopt the standard approach \cite{girardeau,gardiner,castin}.  
We use the identity $\hat{a}_0 ^{\dagger} 
\hat{a}_0 = \hat{N} - \sum ^{'} \hat{a}_{\v k}^{\dagger} 
\hat{a}_{\v k}$ to eliminate the zero mode number operator (a 
prime indicates the omission of the zero mode). 
 In a large $N$ approximation we keep terms up to second order in $\hat{a}_{\v k}^{\dagger}$ and $\hat{a}_{\v k}$. 
The Hamiltonian then commutes with $\hat{N}$ and choosing an 
appropriate basis replaces $\hat{N}$ by its eigenvalue $N$.  
$\hat{a}_{0}^{\dagger}$ and $\hat{a}_{0}$ can be replaced by 
$\sqrt{\hat{N}_0} = \sqrt{\hat{N} - \sum ^{'} \hat{a}_{\v k}^{\dagger}  \hat{a}_{\v k}}\simeq \sqrt{N}$ \cite{comment2}. 

In order to diagonalize the Hamiltonian (\ref{H2}) we first implement the Bogoliubov transformation in the same approximation. 
The required canonical transformation $U_B=\exp {\bm (}i \hat{G}_b(\theta){\bm )}$ is generated by 
\begin{equation}
\hat{G}_b (\theta) = \frac i2 \sideset{}{'}\sum _{\v k} \theta _{k}  \hat{a}_{\v k}^{\dagger}  \hat{a}_{-\v k}^{\dagger}+{\rm h.c.},
\end{equation}
with 
\begin{equation}
 \theta _{k} = \tanh ^{-1}\left(\frac{gn}{\hbar \omega _{k} +\epsilon _{k}+gn}\right), 
 \ \hbar \omega _{k} = \sqrt{\epsilon _{k}(\epsilon _{k}+2gn)} .
\end{equation}
Bogoliubov's excitation (the ``bogolon") is created and annihilated by 
\begin{align}
\hat{b}_{\v k}^{\dagger} & =
e^{i \hat{G}_b (\theta)} \hat{a}_{\v k} ^{\dagger}\;e^{- i \hat{G}_b (\theta)}
= \hat{a}_{\v k} ^{\dagger} \;\cosh \theta _{k} +\hat{a}_{-\v k}\;\sinh \theta _{k}, \\
\hat{b}_{\v k}& =
e^{i \hat{G}_b (\theta)} \hat{a}_{\v k} \;e^{- i \hat{G}_b (\theta)}
=\hat{a}_{\v k} \;\cosh \theta _{k} + \hat{a}_{-\v k}^{\dagger}\;\sinh \theta _{k} .
\end{align}
The original vacuum state $| 0 \rangle$ is transformed to the new bogolon vacuum: 
\begin{align}
| 0 (\theta) \rangle & = e^{i \hat{G}_b (\theta)} | 0 \rangle \\ \nonumber
&= \sideset{}{'} \prod _{\v k} \frac{1}{\cosh \theta _{k}} 
\exp\left(\hat{a}_{\v k} ^{\dagger}\hat{a} _{- \v k} ^{\dagger}\;\tanh \theta _{k}\right) | 0 \rangle .
\end{align}
The Hamiltonian after this Bogoliubov transformation is 
\begin{equation} \label{H5}
\hat{H}_B(t)= E_0
+\sideset{}{'} \sum _{\v k} \hbar \omega _{k} \hat{b}_{\v k}^{\dagger}  \hat{b}_{\v k}
+ \sideset{}{'} \sum _{\v k} {\bm (}f_{\v k}(t) \hat{b}_{\v k}^{\dagger}+{\rm h.c.}{\bm )} ,
\end{equation}
where the ground state energy $E_0$ with no impurities is 
$E_0 =  g n^2 V/2 +\sideset{}{'} \sum (\hbar \omega _{k} -\epsilon _{k}-gn)/2 
=  (gn^2V/2) [1+(128/15) \sqrt{n a_s^3/\pi}\;]$, and 
\begin{equation} \label{f}
f_{\v k}(t)= \sqrt{\frac{n\epsilon _{k}}{V\hbar \omega _{k}}} \sum _i  \lambda_i \;e ^{-i\v k\cdot \v{\zeta}_i(t)} .
\end{equation}

The Hamiltonian (\ref{H5}) describes a system of decoupled forced harmonic oscillators and, 
therefore, can be solved exactly \cite{schwinger,gross}. 
Diagonalization of (\ref{H5}) is easily accomplished by way of time dependent unitary transformation 
$U_C=\exp{\bm (}i\hat{G}_c(t){\bm )}$ generated by 
\begin{equation}
\hat{G}_c(t) = i \sideset{}{'}\sum _{\v k} \phi _{\v k}(t)  \hat{b}_{\v k}^{\dagger} +{\rm h.c.}
\end{equation}
This transformation shifts $ \hat{b}_{\v k}^{\dagger}$ and $ \hat{b}_{\v k}$: 
\begin{align} \label{c}
e^{i \hat{G}_c(t) }\hat{b}_{\v k}^{\dagger} \;e^{- i \hat{G}_c(t)}
&= \hat{b}_{\v k} ^{\dagger}+  \phi _{\v k}^{*}(t), \\
e^{i \hat{G}_c(t) }\hat{b}_{\v k} \;e^{- i \hat{G}_c(t)}
&= \hat{b}_{\v k} +  \phi _{\v k}(t) .
\end{align} 
In order for $\hat{G}_c(t)$ to diagonalize the transformed Hamiltonian 
in the Schr\"odinger picture, 
\begin{equation} \label{hc}
\hat{H}_C(t) = e^{i \hat{G}_c(t) } \hat{H}_B(t) \;e^{- i \hat{G}_c(t)} 
-i \hbar \;e^{i \hat{G}_c(t) } \pd _t \;e^{- i \hat{G}_c(t)} , 
\end{equation}
$\phi _{\v k}(t)$ should satisfy a simple differential equation: 
\begin{equation}
i \hbar \pd _t \phi _{\v k}(t) = \hbar \omega _{k}\phi _{\v k}(t) + f_{\v k}(t) .
\end{equation}
We solve this differential equation using the initial condition $\phi _{\v k}(t_0)=0$.  
It is assumed that the interaction is switched on at $t=t_0$.  
\begin{equation} \label{phi}
\phi _{\v k}(t)  =  - \frac{i}{\hbar} \sqrt{\frac{n\epsilon _{k}}{V\hbar \omega _{k}}} \;e^{-i \omega _{k} t} \sum _i  \lambda_i
\int ^t _{t_0} dt' \ e^{i \omega _{k} t' -i\v k\cdot \v{\zeta}_i(t')} .
\end{equation}
After the unitary transformation is performed we find that the Hamiltonian (\ref{hc}) is diagonalized as 
\begin{equation} \label{H6}
\hat{H}_C(t)= \tilde{E}_0 (t) +\sideset{}{'} \sum _{\v k} \hbar \omega _{k} \hat{b}_{\v k}^{\dagger}  \hat{b}_{\v k} .
\end{equation}
The energy spectrum $\hbar \omega _{k}$ remains the same as bogolons, 
the gapless excitations characterized by $\omega _{k} \simeq kc$ for a small $k$, 
where $c=\sqrt{gn/M}$ is the speed of sound. 
In the Hamiltonian (\ref{H6}) we have defined a new ground state energy of the system by 
\begin{equation} \label{E0}
\tilde{E}_0 (t) = E_0 +\frac 12 \sideset{}{'} \sum _{\v k}{\bm (}f_{\v k}^*(t)\phi _{\v k}(t)+{\rm c.c.}{\bm )}, 
\end{equation}
with the second term representing the effects of impurities. 
The bogolon vacuum is shifted corresponding to the unitary transformation $U_C$ as  
\begin{align}\label{vacuum}
| \tilde{0}(t) \rangle &= e^{i \hat{G}_c(t)} | 0 (\theta) \rangle \\ \nonumber
& = \exp [-\sideset{}{'} \sum _{\v k} {\bm (}\;\frac 12 | \phi _{\v k}(t) |^2+\phi _{\v k}(t) \hat{b}_{\v k}^{\dagger}{\bm )} ]| 0 (\theta) \rangle .
\end{align}

The physical interpretation of the vacuum (\ref{vacuum}) is quite simple. 
The motion of impurities creates time dependent coherent states. 
The new vacuum describes the deformed condensate 
and contains infinitely many bogolons \cite{aft}. 

We next proceed to evaluate the occupation number 
$n_{\v k}$ from which various physical quantities such as the 
dissipated energy, the depletion of the condensate, etc. can be derived.

The expectation value of the number operator $\hat{n}_{\v k}= \hat{b}_{\v k}^{\dagger}  \hat{b}_{\v k}$ 
with respect to the shifted original vacuum $\exp{\bm (}i \hat{G}_c(t){\bm )} |0 \rangle$ is 
\begin{equation}
n_{\v k}=\langle 0| e^{-i \hat{G}_c}\; \hat{b}_{\v k}^{\dagger}  \hat{b}_{\v k} \;e^{i \hat{G}_c} |0\rangle
= \frac{\hbar \omega _{k} +\epsilon _{k} +gn}{2 \hbar \omega _{k}} + \tilde{n} _{\v k} (t) ,
\end{equation}
where $\tilde{n} _{\v k} (t)=| \phi _{\v k}(t) |^2$ is the contribution to $n_{\v k}$ due to the motion of impurities. 
From Eq. (\ref{phi}) we find 
\begin{equation} \label{n}
\tilde{n} _{\v k} (t) = \frac{n\epsilon _{k}}{\hbar^{3} V\omega _{k}} \ |\sum _i  \lambda_i I_i(t) |^2 ,
\end{equation}
where $I_i(t)$ is defined by 
\begin{equation} \label{I}
I_i(t)= \int _{t_0} ^t dt' \ e^{i \omega _{k} t'- i \v k\cdot \v{\zeta}_i(t')} .
\end{equation}

The energy ${\cal E}_{\v k} (t) $ dissipated due to the motion of impurities in a given mode $\v k$ 
is calculated as the expectation value 
of the occupation number with respect to the new vacuum (\ref{vacuum}) multiplied 
by the exciation energy $\hbar \omega _{k}$,  
\begin{equation}
{\cal E}_{\v k} (t) = \langle \tilde{0}(t)|  \hbar \omega _{k} \hat{b}_{\v k}^{\dagger}  \hat{b}_{\v k} | \tilde{0}(t) \rangle  .
\end{equation}
The total dissipated energy ${\cal E}_{{\rm tot}} (t) $ is the sum over all modes $\v k$: 
\begin{equation}
{\cal E}_{{\rm tot}} (t) =  \sideset{}{'} \sum _{\v k} \hbar \omega _{k} \tilde{n}_{\v k} (t)
=  \frac{n}{\hbar^{2} V} \sideset{}{'} \sum _{\v k} \epsilon _{k} \ |  \sum _i \lambda_i I_i(t) |^2 . \label{dissipation}
\end{equation}


The depletion of the condensate is defined by 
\begin{equation}
d(t)= \frac 1N  \sideset{}{'} \sum _{\v k} \langle \tilde{0}(t)| \hat{a}_{\v k}^{\dagger} \hat{a}_{\v k} | \tilde{0}(t) \rangle .
\end{equation}
The straightforward computation gives the result   
\begin{multline}
d(t) =  \frac 83 \sqrt{\frac{n a_s^3}{\pi}} + \frac 1N  \sideset{}{'} \sum _{\v k} \frac{1}{\hbar \omega _{k}}[ (\epsilon _{k} +gn) \tilde{n} _{\v k} (t) \\
-  gn \;{\rm Re} {\bm (}\phi _{\v k}(t) \phi_{-\v k}(t) {\bm )} ].
\end{multline}
In the thermodynamic limit the second term is of the order of $1/N$ and will not contribute to the depletion. 

To illustrate our model consider the simplest case of one classical impurity moving with a constant velocity 
$\v v=v \hat{\v z}$ in the $z$-direction. 
For concreteness assume that the particle starts moving from one end of the box ($z=0$) at $t=0$ 
and arrives at the other end of the box ($z=L=V^{1/3}$) at $t=t_f=L/v$. 
Our main objective is to calculate the occupation number $\tilde{n}  _{\v k} (t)$ in the thermodynamic limit of large $t$.
Therefore, we need to compute $I_{\infty} = \lim _{t_f\to \infty} I(t_f)$. 
We then face a formal mathematical difficulty because $| I_{\infty}|^2$ is the square of the distribution:  
\begin{equation}
I_{\infty} =  \lim _{\eta \to 0^+} \int _0 ^{\infty} dt \ e^{iw t -\eta t}
= {\cal P} (\frac{i}{w}) +\pi \;\delta(w) ,
\end{equation}
where $\cal P$ is the principal value distribution, $\delta$ is the Dirac distribution, and $w=\omega _{k} - \v k\cdot \v v$. 
The standard way to deal with this difficulty is as follows, 
\begin{equation} \label{Iv}
|I(t) |^2 = 2 \frac{1-\cos (wt)}{w^2} \simeq 2 \pi t \;\delta(w) ,
\end{equation}
where $t$ is large. 
Thus $|I(t) |^2$ is proportional to $t$ with a coefficient that 
becomes a delta function in the limit $t \to \infty$. Lastly, we remark that the 
replacement $|I(t)|^2/t \to 2 \pi \;\delta (w)$ is valid only asymptotically and the 
actual peak always has a finite width. 

We find that the rate of energy dissipation in the thermodynamic limit is 
\begin{equation} \label{de}
\gamma \equiv \frac{d {\cal E}_{{\rm tot}}(t)}{dt}
= \frac{ n \lambda^2}{(2\pi \hbar)^2} \int d^3 k\;\epsilon _{k} \;\delta(\omega _{k}  -  \v k\cdot \v v).
\end{equation}
Thus $\gamma=0$ unless there are such values of $\v k$ for which $\omega _{k}  =  \v k\cdot \v v$. 
To proceed further, we rewrite (\ref{de}) in spherical coordinates and introduce new variables 
$ x= \hbar k/(2Mc)$, $y= \beta \cos \theta $, where $\beta=v/c$. 
The result of evaluation of the integral (\ref{de}) is 
\begin{equation}\label{de2}
\gamma =  \frac{4n \lambda^2M^3c^3}{\pi \hbar^4 \beta} \int_0^{\infty} dx\ x^3 \int _{-\beta} ^{\beta} 
dy\ \delta( \sqrt{x^2+1}-  y).
\end{equation}
Clearly it follows from Eq. (\ref{de2}) that $\beta$ has to exceed 1 in order to have energy dissipation 
and hence, Landau's criterion is derived. 
The rate of energy dissipation is: 
\begin{equation} \label{gamma}
\gamma = \frac{n \lambda^2M^3c^3}{ \pi \hbar^4 \beta} (\beta^2-1)^2 \;\Theta(\beta-1) ,
\end{equation}
where $\Theta (x)$ the step function. 
Like Cherenkov radiation this radiation is emitted inside a cone with opening angle $ \theta _c = \cos ^{-1} (1/ \beta)$, 
and its momentum cuts off at $\hbar k_c=2Mc\sqrt{\beta^2 -1}$ keeping the total dissipated energy finite. 
Our result agrees with that obtained previously in another approaches based on the 
Gross-Pitaevskii equation \cite{kovrizhin, ms, astracharchik}. 

As another example consider a single classical impurity moving with 
a constant acceleration $a$ in the $z$-direction. We adopt the 
relativistic form of the trajectory $\v{\zeta} (t)=(c^2/a)\sqrt{1+(at/c)^2} \hat{\v z}$ 
(which reduces to $ a t^2/2 +$constant in the non-relativistic limit) 
so that the velocity of the particle is always bounded by the speed of sound $|\dot{\zeta}(t)| < c$. 
To obtain analytic results we extend the range of integration in (\ref{I}) to $(- \infty, +\infty)$, taking 
into account the complete trajectory. This corresponds to an 
idealized thermodynamic limit. In reality we always have to use 
finite times, volumes, and trajectories conforming to the experimental situations.

The integral (\ref{I}) is now evaluated, after the change of variables from $t$ to $\tau$, $t=( c/a) \;\sinh \tau$,
\begin{align}
I_{\infty} &= \frac{c}{a} \int _{-\infty}^{\infty} d \tau\  e^{i \Omega \;\sinh (\tau - \chi)} \;\cosh \tau  \\ \label{Iacc}
&= \frac{2c}{a} {\bm (} \pi \delta (\Omega)\;\cosh \chi + i {\rm K}_1 (\Omega)\;\sinh \chi  {\bm )},
\end{align}
where $\Omega =(c/a) \sqrt{\omega_k ^2- (c \v k \cdot \hat{\v z})^2}$, 
$\chi = \tanh ^{-1} (c \v k \cdot \hat {\v z}/\omega_k)$, 
and ${\rm K}_1 (x)$ is a modified Bessel function of the second kind. 
Since $\Omega =0$ if and only if $k=0$, the 
first term in Eq. (\ref{Iacc}) does not contribute to the occupation number after 
multiplying by $\epsilon _k/\omega _k $ in Eq. (\ref{n}) and using $x\;\delta (x)=0$.  
Hence we obtain 
\begin{equation} \label{nonthermal}
\tilde{n}_{\v k}= \frac{4 c^2 n \lambda ^2\epsilon _k}{V\hbar^3 a^2 \omega _k}  {\bm (} {\rm K}_1 (\Omega) \;\sinh \chi{\bm )}^2  .
\end{equation}
In contrast to the constant velocity case there is an exponential tail at high 
momenta (which follows from the asymptotic behavior of ${\rm K}_1 (x)$ for $x\to \infty$),
\begin{equation} \label{nacc}
\tilde{n}_{\v k} \simeq \frac{16 \pi n\lambda ^2 M^3 c^3 \;\cos ^2 \theta}{V \hbar^5 a k^4} \;\exp\left(-\frac{\hbar c k^2}{ M a}\right),
\end{equation}
where $\theta$ is an angle between the directions of the momentum $\hbar\v k$ and the trajectory of the impurity. 
The formula (\ref{nacc}) displays the exponential dependence on the mass $M$ of a heavy boson. 
The particle count by a detector can be used to measure $M$. 
The total dissipated energy is evaluated in the weak acceleration limit by using again the asymptotic form of ${\rm K}_1 (x)$,  
\begin{equation}
{\cal E}_{{\rm tot}} \simeq \frac{ n \lambda ^2 M^2 c}{10 \hbar ^3} \sqrt{\frac{\hbar a}{\pi M c^3}} .
\end{equation}
Thus we find that although the velocity of the impurity never exceeds the speed of sound, 
there is a finite amount of energy dissipated due to the motion of impurity. 
This result may appear couterintuitive. 
However this is not an artifact due to the presence of an infinite time interval in our calculation. It simply means that 
Landau's phenomenological argument based on the Galilean covariance does not apply to the constant acceleration case.


As we have seen, our simple model Hamiltonian (\ref{H2}) yields 
Landau's criterion in the case where the impurity moves with a 
constant velocity. In contrast to the original argument, however, 
we find finite energy dissipation even for slightly sub-critical velocity $v$. 
This can be seen from Eqs. (\ref{Iv}--\ref{gamma}).

We have also obtained the dissipated energy for a constantly accelerated impurity in the weak acceleration limit. 
In order to estimate experimental dissipation for this case and the previous case, 
we should evaluate expressions like (\ref{dissipation}) in a way that corresponds to the actual numerical parameters. 

Our model is applicable to more general cases such as several impurities moving with 
different velocities or in a completely random way. Our model should be also extended to finite temperatures 
in order to deal with real experiments. 

Finally, we point out that the exact expression for the occupation 
number (\ref{nonthermal}) shows that the vacuum condensate 
does not mimic a simple thermal bath usually associated with the accelerated motion of impurities. 
The results of an extended analysis of the latter problem will be presented in the forthcoming publication \cite{ms2}.

\begin{acknowledgments}
The author would like to acknowledge Professor Pawel O. Mazur for suggesting this problem 
and helpful discussions, especially regarding the constant acceleration case.
He also thanks Professor Shmuel Nussinov for useful discussions.
This work was supported in part by the National Science Foundation, Grant No. PHY-0140377.
\end{acknowledgments}  
\noindent
$^*$ e-mail: suzuki@physics.sc.edu



\begin{thebibliography}{9}
\bibitem{comment2} This approximation is valid after multiplying $\hat{a}_{0}^{\dagger}$
or $\hat{a}_{0}$ by $\hat{a}_{\v k}^{\dagger}$ and/or $ \hat{a}_{\v k}$.
\end{thebibliography}
\end{document}